# Scale-invariant jet suppression across the black hole mass scale

David Garofalo[1] and Chandra B. Singh[2]

1. Department of Physics, Kennesaw State University, Marietta GA 30060, USA
2. Department of Astronomy (IAG-USP), University of Sao Paulo, Sao Paulo, Brazil

Abstract

We provide a schematic framework for understanding observations of jet suppression in soft state black hole X-ray binaries based on the Blandford-Payne process and the net magnetic flux threading the black hole. Due to the geometrical thinness of soft state disks, mass-loading of field lines is ineffective compared to both geometrically thick disks as well as thin disks with greater black hole threading flux, a simple physical picture that allows us to understand the weakness of jets in radiatively efficient thin disks accreting in the prograde direction around high-spinning black holes. Despite a simplicity that forbids insights into the complexity of turbulent-driven evolution or the physics of the observed short-term time variability, we show how the breadth of this framework is such that it can serve as a coarse-grained foundation for understanding black hole accretion and jet formation across the mass scale.

1. Introduction

Despite the adoption of scale-free models of black hole accretion and jet formation over the past few decades (Sikora et al 2007; Meier 2001; Moderski et al 1998; Wilson & Colbert 1995; Blandford 1990; for a recent review see Czerny & You 2015), the observations of jets across the mass scale remain difficult to interpret. In particular, while on small scales we observe black hole X-ray binaries without noticeable jets in soft or radiatively efficient accretion states, FRII quasars on the largest scales, instead, appear to be accreting black holes with both powerful jets and radiatively efficient accretion (Hardcastle, Evans & Croston 2007). This seeming violation of scale invariance appears to be resolved in postulating the FRII quasar phenomenon as the large-scale manifestation of the ballistic jet in X-ray binary state transitions. However, this picture is at odds with the observational fact that FRII quasars experience their peak at higher redshift compared to FRI radio galaxies (Garofalo 2013). If we take the observations of apparent radiatively efficient accretion in FRII quasars seriously, the question that naturally emerges is this: Is there a way of achieving powerful jets in radiatively efficient accretion states and why would this only appear to happen in the large-scale family of accreting black holes? Our goal in this work is to answer this question based on longstanding physical ideas about accretion and jets, i.e. ideas that are not new. As we will show, the answer lies in understanding the so-called 'jet suppression' mechanism observed in X-ray binaries (Neilsen & Lee 2009; Ponti et al 2012), and in recognizing its scale-invariant character. In this paper we explore a simple phenomenological explanation that nonetheless explains a host of observed phenomena across the mass scale. In section 2 we explore jet suppression in the stellar mass black hole case; in section 3 we extend the phenomenology to the active galaxy population; and in section 4 we conclude.

2. Jets during X-ray binary state transitions

The piece of physics necessary to understand the schematic presented in this work is due to Blandford & Payne (1982), who solved the magnetohydrodynamic (MHD) equations for a self-similar magnetic field configuration. Their conclusion is that the magnetic field must bend sufficiently at the disk surface in order to trap plasma on the field and produce a centrifugally launched wind. Because magnetic pressure due to adjacent magnetic field lines will push and make field lines bend, Garofalo (2009a) showed that a greater black hole-threading magnetic flux induces a larger bend in the magnetic field lines threading the accretion disk. This is shown in Figure 1 with the colored arrow indicating the pressure on the field threading the disk from the field threading the black hole.

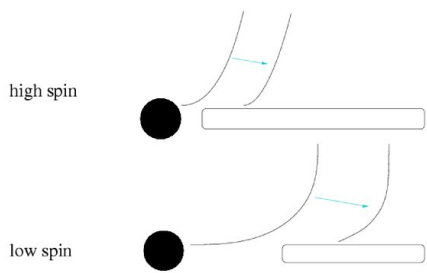

Figure 1: Schematic of the numerical result of Garofalo 2009a whereby a high spinning black hole in a prograde configuration does not accumulate large magnetic flux on the black hole and therefore experiences less bend in the magnetic field lines threading the disk (top). For lower spinning black holes in prograde configurations (or for retrograde configurations), the gap region is larger which leads to larger magnetic flux on the black hole, which produces a greater bend in the disk-threading magnetic field. From Garofalo 2009b.

We appeal to thin vs thick disk geometry and Figures 1, 2, and 3 in order to describe the appearance and disappearance of jets in black hole X-ray binaries. Let us begin with a high spinning black hole surrounded by a thick disk in a prograde configuration (Figure 2) and attempt an explanation of the hard state and its observational association with a mildly relativistic jet (Fender et al 2004). The presence of the jet comes from a combination of two effects, the existence of a rotating black hole with a non-zero spin value (the Blandford-Znajek effect; Blandford & Znajek 1977) and the existence of magnetic fields trapping disk plasma and producing a magnetic-based, centrifugally-driven, outflow (the Blandford-Payne mechanism; Blandford & Payne 1982). Because the disk is geometrically thick as shown in Figure 2, plasma can be easily trapped onto magnetic field lines to produce an effective Blandford-Payne jet that serves to collimate and accelerate the black hole jet (Bogovalov & Tsinganos 2005; Garofalo, Evans & Sambruna 2010). The emphasis is on two components for jet production, namely the black hole jet implying non-zero black hole spin, and plasma trapped effectively on magnetic field lines threading the disk for an effective Blandford-Payne jet.

In Figure 3 we have a schematic of a thin accretion disk, which models the soft state in X-ray binaries (Fender et al 2004). The only difference with respect to Figure 2 is the geometrical thickness of the disk. The magnetic field line, in fact, maintains the same bend at the disk equatorial plane as a result of the fact that it is the value of black hole spin that determines the magnetic flux on the black hole (Garofalo 2009a), which in turn affects the field bending at the disk as shown in Figure 1. While the magnetic flux on the black hole increases with increase in disk thickness, it also decreases with decreasing accretion rate, which means that at least at zeroth-order, the disk thickness does not influence jet power (Reynolds et al 2006; Garofalo 2009a,b). Given how our simple schematic prescribes that the field bend at the disk plane remains unchanged in the state transition, the determining factor in the Blandford-Payne jet is the disk geometry, not the geometry or strength of the magnetic field. As can be seen in Figure 3, the plasma is constrained to live predominantly in the thin disk, quite close to the equatorial plane, unlike the plasma of Figure 2. Hence, less material is captured onto the magnetic field lines. Based only on the geometry, therefore, Blandford-Payne outflows in soft states are less effective, and jets can ultimately be suppressed. Despite its qualitative character, this conclusion has far-reaching implications across the mass scale, as we illustrate.

Appealing to observations of the hard to soft transition, we apply this simple picture to construct a phenomenological explanation of the observational fact of a transitory burst, or momentary jet that is more powerful than in the hard state (Fender et al 2004; Narayan & McClintock 2012). Because the disk thickness decreases during this transition, there is a relatively short time during which plasma lingers on magnetic field lines as the system emerges from the hard state and the radiatively-driven disk wind begins to increase as the thickness of the disk collapses. Hence, there is a brief time during which the collimation of the Blandford-Payne jet is more effective than in the hard state, as a result of a disk wind that was not present during the hard state, but progressively grows in strength as the system transitions into a thin disk. But the plasma eventually abandons the magnetic field and is not resupplied which means the Blandford-Payne jet is suppressed as the system enters the geometrically thin state. And, finally, as the system transitions away from the soft state at lower luminosity and goes back into the hard X-ray state, the increase in disk thickness allows the Blandford-Payne jet to again become effective and a complete cycle results.

It is important to emphasize that the cycle just described applies to a high spin black hole surrounded by a prograde-accreting disk. If we assume the value of the spin to be lower or intermediate, the evolution of the system based on the physics just applied is different. In fact, when the black hole spin is lower, the value of the radius of the innermost stable circular orbit increases and the region between the location of the innermost stable circular orbit and the event horizon radius becomes larger, the so-called 'gap region'. Since this gap region is thought to be instrumental in the amount of magnetic flux that is brought to and threads the black hole (Reynolds et al 2006; Garofalo 2009a), the bending of the magnetic field lines at the disk plane will be different for different black hole spin as shown in Figure 1. Because the gap region is largest for largest retrograde spin, the bending at the disk plane will be greatest at largest retrograde spin. Therefore, everything else being equal, for prograde accreting disks around intermediate spin black holes, the Blandford-Payne jet is more effective compared to the equivalent high spin case, which means that soft state X-ray binaries around intermediate spinning black holes will have jets that are less suppressed compared to the high spin case described above.

Here we have not specified whether the large scale magnetic field is passively advected inwards, produced from dynamo generated disk magnetic field, or some combination of the two processes. However, we briefly discuss previous work arguing that disk dynamos produce large-scale fields in a way that is compatible with our schematic in this paper. Livio (2001) discusses phenomenology associated with the idea that the vertical large-scale magnetic field ($B_z$) can be generated locally from dynamo-generated accretion-disk magnetic field ($B_{disc}$) through some reconnection process. Based on an order-of-magnitude estimate, Pringle (1993) presented the minimum requirement for this as $B_z^{2} / B_{disc}^{2} \sim \dot M_{jet} / \dot M_{disc}$ H/R, where $\dot M_{jet}$ and $\dot M_{disc}$ are mass flux in the jet and accretion disk respectively, and H the disk half-thickness. From observations, H/R and $\dot M_{jet} / \dot M_{acc}$ are estimated in the range 0.03-0.3 and 0.01-0.3, respectively. Assuming that $B_{z}$ is generated by $B_{disc}$ loops having length distribution represented by $n(l) \sim l^{-\delta}$, Livio (2001) showed that \delta should lie in the range 1.7 – 3.4. In another model-dependent exploration, Tout and Pringle (1996) obtained \delta = 2 which lies in the range proposed by Livio (2001). Since $B_{z}$ and $\dot M_{jet}$ seem to depend on H (Livio 2001), the ratio of $B_{z}$ and $B_{disc}$ due to reconnection increases with increase in thickness of the disk and there is a larger value of $\dot M_{jet}$ with respect to $\dot M_{acc}$ in a thicker disk.

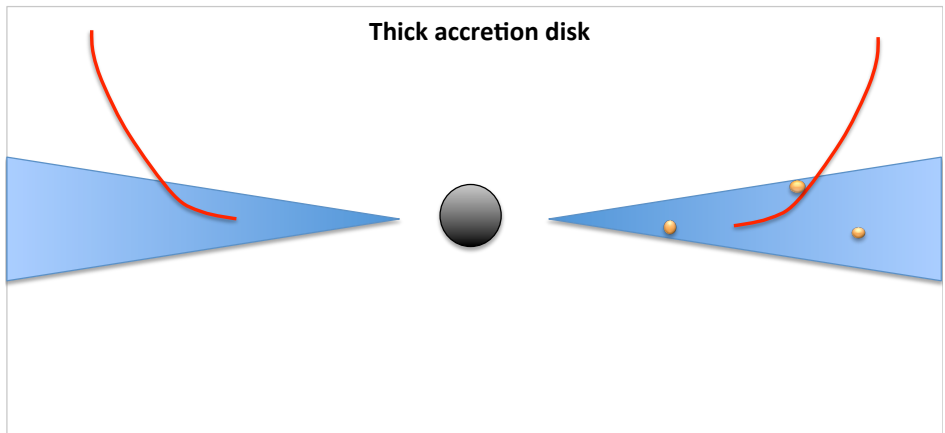

Figure 2: A thick-disk around a high-spinning, prograde rotating, black hole. Note that mass-loading onto magnetic field lines threading the disk can occur at heights that are considerably above the equatorial plane of the disk.

In the next section we apply these simple ideas to active galactic nuclei (AGN) showing how it is possible to appreciate the existence of powerful, collimated jets, despite an association with accretion states that appear to be equivalent to those in soft state X-ray binaries.

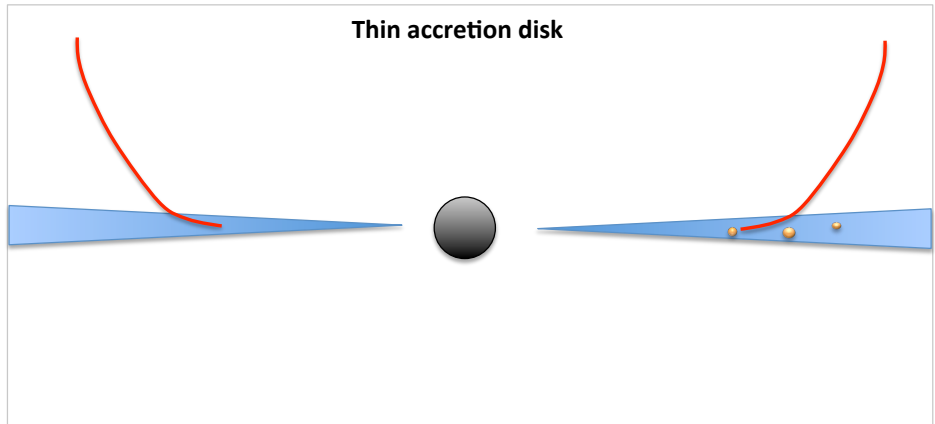

Figure 3: A thin-disk around a high spinning prograde rotating black hole. Note how mass-loading onto magnetic field lines threading the disk occurs only close to the equatorial plane of the disk.

### 3. Jets in AGN

In Figure 4 we show the geometry of the magnetic field threading the disk for a high spinning black hole that is surrounded by a thin accretion disk that rotates in the retrograde configuration. Because the magnetic flux threading the retrograde accreting black hole is larger than in the prograde case (Piotrovich et al 2015), magnetic pressure forces the disk-threading magnetic field lines to bend more toward the equatorial plane (Garofalo 2009a,b), allowing the Blandford-Payne jet to be most effective. Given this understanding, it is now possible to appreciate the absence of jet suppression in retrograde accreting black holes. Despite the thinness of the disk, magnetic fields are significantly bent at the disk equatorial plane as a result of the back-reaction of the black hole threading magnetic flux on the magnetic field threading the accretion disk (Garofalo 2009b). Hence, the powerful FRII quasars which are modeled as high spin retrograde accreting black holes (Garofalo, Evans & Sambruna, 2010; Gnedin, Mikhailov & Piotrovich, 2015), are understood to produce the most powerful and most collimated jets (due to the largest disk threading and hole threading fields), in tandem with a radiatively efficient accretion mode (a thin disk), the largescale equivalent of the soft states as the schematic of Figure 3 illustrates. From the picture we have just presented, it also follows that thick disks around retrograde accreting black holes produce effective jets as described in Garofalo, Evans & Sambruna, 2010, and that according to this framework, FRI radio galaxies are the supermassive black hole analogs of the hard state X-ray binaries. These ideas also allow for an understanding of the greater bend for the retrograde case and how this is the reason why the jet in the hard state for X ray binaries, and the jet in FRI radio galaxies are weaker than both the ballistic jet and the FRII quasar jet, respectively. In other words, hard state jets in X-ray binaries and FRI radio galaxy jets are modeled as prograde accreting systems that have weaker flux bundles on their black holes, thereby producing weaker bending in the magnetic fields threading their disks.

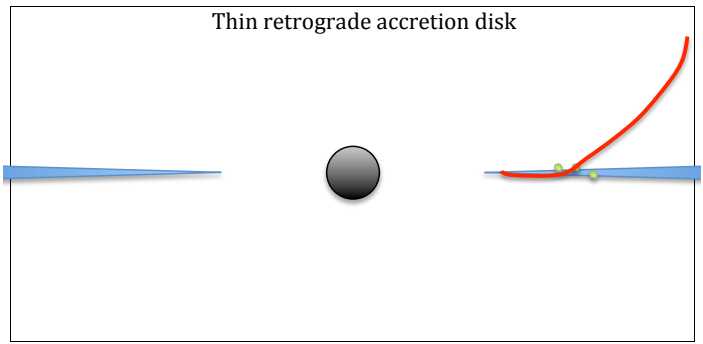

Figure 4: A thin-disk around a high spinning retrograde accreting black hole. Because the innermost stable circular orbit is further away from the black hole, the black hole threading flux is larger than in the prograde case, and a larger flux bundle on the hole produces larger magnetic pressure on the disk-threading field. As a consequence, the disk-threading field bends more toward the equatorial plane (Garofalo 2009b). And the large bend in the magnetic field line threading the disk allows for effective mass-loading and an effective Blandford-Payne jet.

## 4. Conclusions

In this paper we have produced a simple schematic picture of basic physics about accretion and jets developed decades ago that have been combined into a framework at allows one to produce a contradiction-free, scale-invariant, picture for the behavior of jets and their observed suppression in X-ray binary soft states. The simplicity of the picture is surely not meant to be representative of the details of the accretion physics, which is a complicated chaos due to turbulence, but merely to serve as a zeroth-order description, and is not therefore expected to explain observed processes such as short-term time variability in accreting black holes. But a wealth of observed features related to black hole accretion fit within this scenario; and, fundamentally, it accommodates scale invariance. While on the stellar-mass black hole scale it explains general features of the evolution from hard states to soft states associated with jet suppression, the mechanism allows one to appreciate the time evolution of soft state-accreting black holes with powerful jets toward less effective jet states as black holes spin up toward the prograde regime in the supermassive black hole case, namely the radio quiet quasar family of AGN. Because retrograde accretion is difficult to generate in donor star feeding, the small-scale equivalent of the FRII quasar is hard to produce in black hole X-ray binaries, and has therefore most likely never been observed.

It is also worth re-emphasizing that to zeroth-order in our theoretical framework, the jet suppression mechanism described here is not directly dependent on the accretion state, whether thick or thin disk. Rather, it depends on the size of the gap region, which in turn depends on the value of black hole spin (high vs low) and the orientation of the disk (retrograde vs prograde). The more magnetic flux is accumulated on the black hole, the more the disk-threading magnetic field bends, and the greater the ability to produce a Blandford-Payne jet. However, if the disk-threading magnetic field is not sufficiently bent toward the disk, as is the case in high prograde systems, increased disk thickness (such as that in advection dominated disks) will come to the rescue by essentially bringing plasma to the magnetic field at heights that allow for the generation of a centrifugally-driven outflow, effectively allowing the system to behave as if the bending were greater at the disk surface. Note that our picture is not incompatible with the suggestion by Neilsen & Lee (2009) that the radiatively-driven disk wind encompasses a sufficiently large amount of material that an insufficient amount is left for producing the jet. In fact, such a process could add to the weakness of mass-loading in prograde configurations around thin disks.  While the observation of jet suppression was incorporated in models of accretion and jet formation from black holes as a purely observational fact (Garofalo, Evans & Sambruna 2010; Garofalo et al 2016), this is the first time that a physical explanation has been identified within the paradigm. In closing, we note the following equivalence classes: X-ray binary soft states have large scale counterparts in radio quiet quasars or radio quiet AGN.  Hard state X-ray binaries have large scale counterpart in FRI radio galaxies or FRI jet morphologies in low excitation radio galaxies. FRII radio galaxies and FRII quasars are not seen in X-ray binaries due to the fact that the paradigm is founded on the idea that retrograde accretion is unlikely to occur in stellar-mass black hole systems. And, finally, the ballistic jet in X-ray binaries is not observed in the supermassive black hole case due to the fact that AGN accretion does not naturally evolve from a low excitation state to a high excitation one.  In other words, because black holes on different scales feed differently (donor star in X-ray binaries and mergers and secular processes in active galaxies), there is a scale-invariance breaking resulting from different initial conditions.

Acknowledgments

We thank the referee for highlighting ideas in the literature that helped constrain our phenomenology.

References

Blandford, R.D., in Active Galactic Nuclei, ed. By T.J.-L. Courvosier, M. Mayor (Springer, Berlin, 1990), p. 161

Blandford, R.D., & Znajek, R., 1977, MNRAS, 179, 433

Blandford, R.D., & Payne, D. G., 1982, MNRAS, 199, 883

Bogovalov, S. & Tsinganos, K., 2005, MNRAS, 357, 918

Czerny, B. & You, B., 2015, eprint arXiv: 1507.05852

Fender, R.P. Belloni, T.M. & Gallo, E., 2004, MNRAS, 355, 1105

Garofalo, D., 2009, ApJ, 699, 400 (a)

Garofalo, D., 2009, ApJ, 699, L52 (b)

Garofalo, D., Evans, D.A., & Sambruna, R.M., 2010, MNRAS, 406, 975

Garofalo, D., 2013, AdAst, 213105

Garofalo, D., Kim, M.I., Christian, D.J., Hollingworth, E., Lowery, A. Harmon, M. 2016, ApJ, 817, 170

Gnedin, Yu.N, Mikhailov, A.G., Piotrovich, M. Yu, 2015, AN, 336, 312

Hardcastle, M.J., Evans, D.A., & Croston, J.H., 2007, MNRAS, 376, 1849

Livio, M., 2001, ASP, Conference Series vol. 224

Meier, D.L., 2001, ApJ, 548, L9

Moderski, R., Sikora, M., & Lasota, J.-P., 1998, MNRAS, 301, 142

Narayan, R. & McClintock, J.E., 2012, MNRAS, 419, L69

Neilsen, J. & Lee, J.C., 2009, Nature, 458, 481

Piotrovich, M.Y., Buliga, S.D., Gnedin, Y.N., Natsvlishvili, T.M., Silant'ev, N.A., 2015, Ap&SS, 357, 99

Ponti, G. et al 2012, MNRAS, 422, L11

Pringle, J.E., 1993, in Astrophysical Jets, eds. D. Burgarella, M. Livio & C.P. O´Dea (Cambridge: Cambridge University Press), p. 1

Reynolds, C.S., Garofalo, D., Begelman, M., 2006, ApJ, 651, 1023

Sikora, M., Stawarz, L., Lasota, J-P, 2007, ApJ, 658, 815

Tout, C.A., & Pringle, J.E., 1996, MNRAS, 281, 219

Wilson, A.S. & Colbert, E.J.M., 1995, ApJ, 438, 62